# Observation of the Spin-Seebeck Effect in a Ferromagnetic Semiconductor

C. M. Jaworski[1,] J. Yang[2], S. Mack[3], D. D. Awschalom[3], J. P. Heremans[1,4*], R. C. Myers[2,4*]

1. Department of Mechanical Engineering, The Ohio State University, Columbus, OH
2. Department of Materials Science and Engineering, The Ohio State University, Columbus, OH
3. Center for Spintronics and Quantum Computation, University of California, Santa Barbara, CA
4. Department of Physics, The Ohio State University, Columbus, OH

*e-mail: heremans.1@osu.edu; myers.1079@osu.edu

**The spin-Seebeck effect was recently discovered in a metallic ferromagnet and consists of a thermally generated spin distribution that is electrically measured utilizing the inverse spin Hall effect. Here this effect is reproduced experimentally in a ferromagnetic semiconductor, GaMnAs, which allows for flexible design of the magnetization directions, a larger spin polarization, and measurements across the magnetic phase transition. The spin-Seebeck effect in GaMnAs is observed even in the absence of longitudinal charge transport. The spatial distribution of spin-currents is maintained across electrical breaks highlighting the local nature of the effect, which is therefore ascribed to a thermally induced spin redistribution.**

Reducing the heat generated in traditional electronics is a chief motivation for the development of spin-based electronics, called spintronics[1]. Spin-based transistors that do not strictly rely on the raising or lowering of electrostatic barriers can overcome scaling limits in charge based transistors[2]. Spin transport in semiconductors might also lead to dissipation-less information transfer with pure spin-currents[3]. Despite these thermodynamic advantages expected in spin-based devices and after more than a decade of focused spintronics research, little experimental literature exists on the thermal aspects of spin transport in solids. A recent, and surprising exception was the discovery of the spin-Seebeck effect, reported as a measurement of a redistribution of spins along the length of a sample of permalloy (NiFe) induced by a temperature gradient [4]. This macroscopic spatial distribution of spins is, surprisingly, many orders of magnitude larger than the spin diffusion length. This effect has stimulated a strong interest in thermal aspects of spin transport[5]. From a fundamental standpoint, the spin-Seebeck effect, like the charge-Seebeck effect, allows insight into the electronic structure of materials. In particular, the spin-Seebeck effect provides unique experimental opportunities for probing the spin-polarized density of states of materials.

Here we investigate the generality of the spin-Seebeck effect in other ferromagnetic materials. Ferromagnetic semiconductors, like GaMnAs, exhibit strong magnetic anisotropy and spin polarization, providing a contrast to the electronic structure and equilibrium spin polarization in NiFe. We have carried out measurements of the spin-Seebeck effect in GaMnAs. Similar results are observed in the ferromagnetic metal of MnAs (see Supplementary Information), and will be the object of a subsequent article. Platinum strip contacts provide conversion of local spin-currents into transverse voltage by the inverse spin-Hall effect (ISHE) as well as short out any spurious transverse voltages in GaMnAs that could arise from conventional thermal transport phenomena. The macroscopic spatial distribution of spins due to the spin-Seebeck effect follows roughly a $\sinh(x)$ law. This spatial distribution of spin currents is insensitive to longitudinal charge/spin transport, indicating the distributions occur locally, possibly being mediated by phonon interaction with local moments or spins. We observe the onset of the spin-Seebeck effect below the magnetic phase transition and its variation with magnetization direction. In particular, by engineering the magnetic anisotropy out-of-plane, the spin-Seebeck signal is eliminated, as expected. Lastly, evidence is provided that the spin-Seebeck effect in GaMnAs can be observed using GaMnAs itself as the spin detection layer.

The spin-Seebeck effect was first measured in a thin-film sample of a ferromagnet[4], where an in-plane temperature gradient ($\nabla_x T$) was applied. The geometry of that experiment and ours are identical, shown schematically in Fig. 1a. The magnetization (**M**) and therefore the Fermi level spin polarization vector ($\sigma$) can be controlled using an applied magnetic field (**B**). The sample is equipped with platinum strips, which serve as local spin current detectors: when subjected to $\nabla_x T$, a flux of spins (**J**$_S$) diffuse along the $z$ direction into the strips generating an inverse spin Hall voltage [6], $\frac{V_y}{w}\hat{\mathbf{E}}_{\mathbf{ishe}} = \mathbf{E}_{\mathbf{ishe}} = D_{ishe}(\mathbf{J_S} \times \sigma)$, where $D_{ishe}$ is the spin Hall coefficient of platinum and $w$ is the width of the strip.

The samples tested are Ga$_{1-s}$Mn$_s$As thin films grown on (001) SI-GaAs substrates using stoichiometric low temperature molecular beam epitaxy[7] which results in optimal conductivity, high Curie temperature (T$_C$), and high Mn alloying levels[8]. The un-doped semi-insulating substrates are diamagnetic and contain no transition metals, such as Cr. In GaMnAs, the magnetization direction, and therefore the direction of $\sigma$, can be designed using epitaxial strain to be either in-plane or out-of-plane[9]. This enables us to design control experiments, where the inverse spin Hall effect (ISHE) is absent (**J$_S$** × $\sigma$ = **0**), as well as to measure the contributions to the signal induced by classical thermomagnetic effects, such as the Nernst effect, in the active layers themselves. We have tested several pieces each of four wafers with $x$-axis oriented along either the [1$\bar{1}$0] or [110] crystallographic axes, as shown in Fig. 1b. Additional spin-Seebeck measurements on other samples of GaMnAs and MnAs are included in the Supplementary Information. The samples are parallelepipeds 3-5 mm wide (along $y$) by 10-25 mm long (along $x$). The platinum strips are approximately 20-nm thick and deposited over 1-nm thick Ti adhesion layers. These strips are deposited along the $y$-axis at varying $x$ positions and are approximately 0.25 mm wide. Current-voltage measurements between different strips show that the Pt/Ti/GaMnAs contacts are Ohmic. We heat one side of the sample (x = L/2) and fix the other (x = -L/2) to a heat sink to generate a thermal gradient along *x*. The transverse voltage along a strip, $V_y$, is measured while sweeping $B_x$ through the coercive fields.

In a first set of experiments, we study samples with magnetic easy axes oriented in-plane. Magnetization measurements reveal that [110] is a uniaxial hard axis, while [100], [010] and [1$\bar{1}$0] are easy axes, Fig. 1c and d. This behavior is expected due to the existence of both a uniaxial easy axis along [1$\bar{1}$0] and cubic easy axes along [100] and [010] occurring in heavily alloyed, high-$T_C$ GaMnAs[10]. Figure 1e shows the raw value of $V_y$ measured across the strips while sweeping $B_x$ along the magnetic easy axis [1$\bar{1}$0]. The background voltages and EMF pickup are removed and the signal is centered at zero. Since the charge carrier spins (holes in GaMnAs) are exchange coupled to the local Mn magnetic moments ($\sigma$ // **M**) any transverse spin Hall voltage ($V_y$) due to a local spin current along the *z*-axis (**J$_S$** // *z*) switches sign as the magnetization switches with applied field. Note that the magnitude of the switch is $\Delta V_y$, and is actually twice the voltage generated by ISHE, as M reverses direction. The sign of $\Delta V_y$ switches between the hot and cold sides, revealing a spatial dependence to the sign and magnitude of **J$_S$**. This spatial dependence distinguishes the spin-Seebeck effect from all other known thermomagnetic transport phenomena, such as the conventional thermoelectric power (charge-Seebeck $\alpha_{xx}$) and the transverse Nernst-Ettingshausen effect, $\alpha_{xyz}$[11]. $\alpha_{xx}$ is also measured and does not show any steps at the coercive field. When *x* // [110] the magnetic hard axis, Fig. 1f, a sharp switching of $V_y$ is observed at small field following the magnetization switching along the [100] easy axis (Fig. 1d), which lies 45° off of the applied field. As the magnetic field is further increased, $V_y$ shows an opposite field dependence with a lineshape similar to the hard axis magnetization saturation.

The change in transverse voltage, $\Delta V_y$ as defined in Fig. 1e, is measured at nine positions along the sample with $x // [1\bar{1}0]$ and plotted in Fig. 2a revealing a linear dependence on the applied thermal gradient, $\Delta T_x$. From the slope of the line in Fig. 2a we initially obtain the spin-Seebeck signal $(\Delta V_y/2) / \Delta T_x$. In order to express it as a function of the gradients, we define the spin-Seebeck coefficient as, $S_{xy} \equiv \dfrac{E_y}{\nabla_x T} = \dfrac{L\Delta V_y}{2w\Delta T_x}$, in units of a thermoelectric power. Figure 2b plots the temperature and spatial variation of $S_{xy}$ determined by measuring $V_y$ versus $B_x$ and $\Delta T_x$ in stepped sample temperature increments at various contacts across the sample. The temperature dependence of $S_{xy}$ for individual strip contacts (Fig. 2c) reveal that $S_{xy}$ disappears above $T_C$, but otherwise its temperature-dependence is quite different from that of the magnetization as well as that of the thermoelectric power $\alpha_{xx}$, both shown in Fig. 2d. The positional dependence of $S_{xy}$ at selected temperatures is plotted in Fig. 2e. The spin-Seebeck coefficient roughly tracks a sinh(x) function, though the data points do not display perfect odd symmetry about the mid-point ($x = 0$) of the sample. Normalizing this data by the maximum value of $S_{xy}$ (Fig. 2f) reveals that the spatial distribution approximately maintains its line shape at various temperatures. This spatial dependence is in stark contrast to the thermoelectric power $\alpha_{xx}$, which is independent of $x$.

We directly test for a macroscopic spin/charge current along $x$ by polishing away 0.35-mm wide regions of GaMnAs with sandpaper, thereby severing electrical contact (the 2-point resistance between strip contacts increased from 500 $\Omega$ to over 3 M$\Omega$), Fig. 3. If $S_{xy}$ were induced by a longitudinal spin current ($J_S // x$) or macroscopic spin flux accompanying a flux of the charge carriers, then scratching the sample in half would result in two independent samples, creating a $V_y > 0$ above the scratch and $V_y < 0$ immediately below the scratch. Experimentally, we would then expect hysteresis loops of $V_y$ versus $B_x$ exhibiting steps, $\Delta V_y$, with different sign above and below the scratch. Figure 3a shows hysteresis loops from a strip contact before and after a scratch, exhibiting no qualitative change. The slight offset in coercive field arises from unintentional sample tilt from remounting after scratching the sample. This contact is approximately 0.3-mm distant from the scratch. The spatial dependence of $S_{xy}$ is plotted in Fig. 3b revealing no qualitative change in signal resulting from the scratch. More importantly, the two inner contacts within 0.3 mm of the scratch exhibit no change. The temperature dependence of $S_{xy}$ at each contact, comparing Fig. 2c with Fig. 3c, is likewise unaffected by the presence of the scratch. This demonstrates that the spin-Seebeck signal in GaMnAs does not result from a macroscopic, longitudinal spin-current $J_{Sx}$. We suggest that it originates from a perturbation of the statistical distribution function of the spin-polarized charge carriers induced by the temperature gradient. Since charge/spin carriers cannot cross the scratch, the macroscopic spatial distribution of $S_{xy}$ (Fig. 3b) can only be explained by an interaction insensitive to the scratch, for instance a magnetic dipole coupling across it, and/or thermal coupling through the substrate. Because the sinh(x) dependence of the signal is reminiscent of the length-dependence of magnon-phonon coupling in other magnetic semiconductors [12, 13], we suggest that the spatial distribution of $S_{xy}$ reflects the distribution of phonons in the intact GaAs substrate. This conclusion is also supported by observation that $S_{xy}$ shows no dependence on the overall length of the sample (see Supplementary Information).

To further ascertain the origin of the spin-Seebeck signal, we measure a GaMnAs sample with magnetic easy axis out-of-plane, along [001]. In this geometry, we expect no inverse spin Hall voltage in the platinum strip contacts because $\mathbf{J_S} \parallel \boldsymbol{\sigma}$, however the GaMnAs film is now expected to develop a transverse electric field, $E_y$, due to the transverse Nernst-Ettingshausen effect, which is proportional to the temperature gradient and to the out-of-plane magnetization. In contrast to the spin-Seebeck signal, the Nernst voltage exhibits no spatial dependence. To measure it, we place point contacts using silver epoxy along the length of this sample and tilt it 6° off of the xy-plane, thereby allowing the applied field **B** to flip $M_z$, Fig. 4a. The out-of-plane moment $M_z$ is obtained by growing stressed GaMnAs on relaxed InGaAs, exhibiting easy axis behavior along [001], Fig. 4b. As expected, the point contacts show a transverse Nernst signal, shown in Fig. 4c with **B** multiplied by sin(6°) to obtain $B_{001}$. A temperature dependence of this Nernst effect is included in the Supplementary Information. Importantly, $\Delta V_y$ does not exhibit a difference between the hot and cold ends of the sample, compare red and blue data in Fig. 4c. Strip contacts show no signal (green and orange data) proving that (i) as expected, there is no spin-Hall effect when $\mathbf{J_S} \times \boldsymbol{\sigma} = 0$, and (ii) the strip contacts short out the transverse voltage generated by the Nernst effect in the GaMnAs layer.

Lastly, we repeat the spin-Seebeck measurements with **M** oriented in-plane, but instead of platinum strip contacts, we use point contacts. Unlike the strip contacts, which short out any transverse voltages in GaMnAs, the point contacts directly sample the electric field within GaMnAs. Thus, a transverse voltage induced by other thermo-transport effects, like the planar Nernst effect [11], $\alpha_{xy}$, are also included. Raw voltage traces, shown in Fig. 5a, reveal an *x* dependence similar to the strip contact measurements on the same sample (Fig. 1 & 2), but without a change in sign between the hot and cold ends. This suggests that the signal contains contributions of both the spin-Seebeck effect (*x*-dependent) and the planar Nernst effect (*x*-independent). $\Delta V_y$ is linear in $\Delta T_x$ within the experimental error bars (Fig. 5b), but varies along x. This signal cannot be solely due to the planar Nernst effect since it exhibits an *x* dependence, thus it is a mixture of the spin-Seebeck and planar Nernst effects, $S\alpha_{xy}$, which is measured and normalized in the same manner as $S_{xy}$, previously described. The $T$ dependence of $S\alpha_{xy}$ measured for various contacts is shown in Fig. 5c, and demonstrates that $S\alpha_{xy}$ goes to zero above $T_C$, but exhibits an intermediate behavior between the temperature dependence of the spin-Seebeck effect on the strip contacts (Fig. 2c) and that of the planar Nernst effect, which follows the magnetization (Fig. 2d) [11]. This $S\alpha_{xy}$ also reproduces the results of the scratch test performed on the strip contacts. We attempt to separate the spatially independent component of this mixed signal by averaging $S\alpha_{xy}$ across the sample, which should be proportional to the planar Nernst coefficient ($\alpha_{xy}$). Subtracting this average value reveals the spatially dependent component arising from the spin-Seebeck effect (Fig. 5d).

In the absence of strip contacts, which act as spin-current sensors, detection of a spin-Seebeck signal is unexpected. To explain it, we suggest it arises from GaMnAs acting as its own spin-current transducer. We note that $V_y$ observed in point versus strip contacts are experimentally independent because in the absence of ISHE as in Fig. 4, the strip contacts simply act as electrical shorts to any transverse voltage in the sample. $V_y$ measured in strip contacts with M oriented in-plane (Fig. 1 & 2) therefore originates solely from ISHE in the platinum strips. Similarly, the spin-Seebeck component of the signal in Fig. 5 could be due to a self-ISHE occurring in GaMnAs. We note that a similar transverse voltage was also observed in $Ni_{81}Fe_{19}$ with point contacts [14], though of much smaller magnitude than in platinum strips. Here we observe $S_{xy}$ of similar magnitude in strip and point contacts; this may be possible as GaMnAs has a much higher fraction of spin polarized carriers (>85%)[15] than NiFe (~35%)[16]. A quantitative understanding of this signal is of strong interest since, apart from the data in Fig. 5, the (inverse) spin Hall effect has not been reported in ferromagnetic semiconductors like GaMnAs, but is beyond the scope of this study.

In summary, the spin-Seebeck signal is observed in GaMnAs as well as MnAs using platinum strips or point contacts. It is linear with the temperature gradient and non-existent when the magnetization is out-of-plane, as expected. Unlike any other transport coefficient, the spin-Seebeck effect varies spatially along the length of the sample, approximately following a sinh(x) law. By scratching the sample, we show that no longitudinal macroscopic flux of spin exists suggesting that the spatial dependence is related to phonon transport. The absence of longitudinal spin flux raises the question if spin-Seebeck is subject to Onsager reciprocity relations. Its reciprocal, which would be called a spin-Peltier effect, has not been reported yet; it would imply that a flux of spin-polarized particles accelerated electrically or optically carries heat, as magnons do in a temperature gradient[17]

These results show that the spin-Seebeck effect can be used to generate spin-distributions and local spin currents from thermal gradients in both ferromagnetic semiconductors and metals. However, to utilize this phenomenon for coherent spintronic devices, transport must still take place within the spin diffusion length. The thermodynamics of spintronics, or thermal spintronics, could play an important role in the development of semiconductor spintronic devices given the growing need for energy efficient electronics.

**Methods**

The samples in this study are epitaxial $Ga_{1-s}Mn_sAs$ grown on semi-insulating [001] GaAs substrates using a Veeco Gen II MBE system. The sample presented in Fig. 1,2,3, and 5 was 30-nm GaMnAs grown at 150 °C and with s=0.158. The Mn concentration was calibrated using GaAs and MnAs RHEED oscillations. After growth, the sample was heated to 180 °C for 30 minutes to accomplish an *in-situ* anneal to increase the Curie temperature. This particular sample was grown with substrate rotation to prevent any gradients across the sample, but the As:Ga flux ratio was carefully calibrated using low temperature stoichiometric non-rotated growth calibrations [7], in which the As:Ga ratio and therefore the stoichiometry of the films can be tuned[8]. An additional rotated sample of $Ga_{1-s}Mn_sAs$ (s=0.056) was grown on relaxed InGaAs on a GaAs substrate, data presented in Fig. 4. The relaxed InGaAs layer acts to strain the GaMnAs in tension resulting in an out-of-plane easy axis [001], whereas all other samples exhibited in-plane easy axes. Additional samples were grown and tested as described in the Supplementary Information.

The wafers are cleaved into samples 3-5 mm wide by 10-25 mm long along either [110] or [1$\bar{1}$0] crystal directions. A layer of Ti less than 1-nm thick was deposited onto the GaMnAs for adhesion followed by 20 nm of Pt in an electron beam evaporator through a shadow mask. Current-voltage measurements verified the Ohmic nature of the contacts.

Sample magnetic characterization from 2-300 K was performed in a superconducting quantum interference device (SQUID) magnetometer. The magnetic field was oriented parallel to uniaxial and cubic switching directions. Magnetization hysteresis loops were recorded at temperatures corresponding to spin-Seebeck measurements. We subtracted the diamagnetic background of the GaAs substrate. The samples were measured over the temperature range 40-300K in high vacuum using the Thermal Transport Option (TTO) in a Quantum Design Physical Properties Measurement System. Images of the measurement setup are shown in the Supplementary Information. Cernox thermometers attached to gold plated manganin leads are used to determine longitudinal temperature gradients. We measure with and without the thermometry attached, with no change in signal sign or magnitude. We attached 0.001" diameter copper wire with silver epoxy on the edges of the samples on GaMnAs or on platinum strip contacts. While we exercise great care to minimize and keep constant the size of the contacts, variations in contact size generate error. We step temperature and heater power, and after sufficient stabilization time (~ 1hr) record $\Delta T_x$ and $V_y$ using a Keithley 2182A nanovoltmeter while sweeping magnetic hysteresis loops. We are forced to sweep magnetic field, instead of stepping and averaging $V_y$, in order to remove the effect of thermal drift on background voltages. Field is swept at an average rate of 13 Oe/s, thus each hysteric sweep takes approximately 150 seconds. The trade-off between integration time and sweep rate was dictated by the need to find a minimum in the compromise between noise and drift. We note that differences in the magnetization data and spin-Seebeck data could arise since these measurements were performed in different instruments, using different sweep methods (step for magnetization and sweep for spin-Seebeck), and the voltmeter has a 100 millisecond integration time. Measurements on the strips nominally have an RMS noise of 10-15 nV. A report of zero $S_{xy}$ means that $\Delta V_y$ is less than the noise. Error in this study stems not only from the signal to noise ratio, but also from the thermometry error in $\Delta T_x$. These two effects counteract each other. At lower temperatures, $\Delta V_y$ increases, but $\Delta T_x$ decreases. Indeed, the extremely high conductance of the GaAs substrates makes establishing a sufficient temperature gradient to allow measurement difficult, and the power one can dissipate in the heater is limited by the ability of the cryostat to maintain a stable temperature. The best compromise between these considerations determines not only the error bar, but also the lowest temperature we report data. Geometrical constraints limit us to placing the external magnetic field to within 10° of the xy plane.

Conventional thermopower ($\alpha_{xx}$) is measured using the same setup in continuous mode with a sweep rate of 1 K/m.

**Acknowledgements**

This work was supported by the NSF, NSF-CBET 0754023, the ONR, and the Ohio Eminent Scholar Discretionary Fund. Partial support was provided by The Ohio State University Institute for Materials Research.

**Additional information**

The authors declare no competing financial interests. Supplementary information accompanies this paper online. Correspondence and requests for materials should be addressed to R.C.M. and J.P.H.

**Figure Captions**

**Figure 1. Measurement of spin-Seebeck effect in GaMnAs using strip contacts**. **a**, Measurement geometry (not to scale). **b**, Crystal directions in GaMnAs. **c** and **d**, Magnetization, **M**, as a function of applied magnetic field, **B** oriented along the easy [1$\bar{1}$0], [100], and hard [110] axes. Inset: hysteresis loop along [110] on a larger scale. **e** and **f**, Transverse voltage, $V_y$ as a function of **B**, along the easy [1$\bar{1}$0] and hard [110] axes with an applied $\Delta T_x$ of 1.77 K and 3.13 K, respectively. Data are shown on strips near the hot and cold ends of the sample. Magnetic field in **c-f** is swept in a hysteretic fashion.

**Figure 2. Temperature and spatial dependence of the spin-Seebeck effect.** Data were taken along the magnetic easy axis, x // [1$\bar{1}$0]. **a**, The change in transverse voltage, $\Delta V_y$ as a function of the applied thermal gradient, $\Delta T_x$, for strips along the length of sample. **b**, Temperature and spatial dependence of the spin-Seebeck coefficient, $S_{xy}$. **c**, Linecuts of this data at various sample positions. **d**, Temperature dependence of magnetization, **M**, and charge-Seebeck coefficient, $\alpha_{xx}$. Magnetization was measured while warming the sample in 300 Oe after cooling the sample from room temperature in a 10 kOe field. Lines are drawn to guide the eye. **e**, Spatial dependence of $S_{xy}$ at selected temperatures. Data are replotted in **f** after normalizing by the maximum value of $S_{xy}$ ($|S_{xy}^{max}|$). Lines in **e** and **f** are fits to sinh(x).

**Figure 3. Experimental test for a longitudinal spin-current due to the spin-Seebeck effect.** All measurements taken with x // [1$\bar{1}$0] on a sample cleaved from the same wafer as in Figs. 1 and 2. Inset (not to scale) indicates the position of the scratch, which was sufficient for complete electrical isolation (>3 M$\Omega$). **a,** Transverse voltage, $V_y$ as a function of applied field, **B** from the strip contact 0.3-mm above the scratch (star) with an applied $\Delta T_x$ of 0.63 K. **b**, Spatial

dependence of the spin-Seebeck coefficient, $S_{xy}$ before and after the scratch. The scratched region is indicated by the shaded region. **c**, Temperature dependence of $S_{xy}$ after the scratch at various positions along the sample (see Fig. 2c for data before the scratch). Lines are drawn to guide the eye.

**Figure 4. Measurements with out-of-plane magnetization. a,** Sample layout (not to scale). Strained GaMnAs on InGaAs results in an out-of-plane magnetic easy axis [001]. **b**, Out-of-plane magnetization, **M** as a function of magnetic field, **B**. **c**, Transverse voltage, $V_y$ versus $B_{001}$ measured on point contacts (the transverse Nernst-Ettingshausen effect) and on strip contacts with **B** 6° tilted from the xy-plane. The applied $\Delta T_x$ is 0.63 K and 0.26 K for the point contacts and strip contacts, respectively.

**Figure 5. Measurement of spin-Seebeck effect using point contacts.** All data taken from the same sample as in Figs. 1-3. Point contacts were used as shown in the schematic (not to scale) with x // [1$\bar{1}$0]. **a,** Transverse voltage, $V_y$ as a function of applied magnetic field, **B** at the hot and cold ends of the sample with an applied $\Delta T_x$ of 0.67 K. **b,** The change in transverse voltage, $\Delta V_y$ as a function of the applied thermal gradient, $\Delta T_x$, for strips along the length of sample. The data are a mixture of a planar Nernst effect and a spin-Seebeck signal, $S\alpha_{xy}$, as described in the text. **c**, $S\alpha_{xy}$ as a function of the sample temperature for differently positioned contacts. Inset plots the spatial dependence of $S\alpha_{xy}$ at a selected temperature. Lines in **b** and **c** are drawn to guide the eye. **d,** The average value of $S\alpha_{xy}$ across the sample (proportional to $\alpha_{xy}$) is plotted as circles. The temperature dependence of the spin-Seebeck coefficient, $S_{xy}$, is estimated by subtracting the average value of $S\alpha_{xy}$ from its value at each contact.

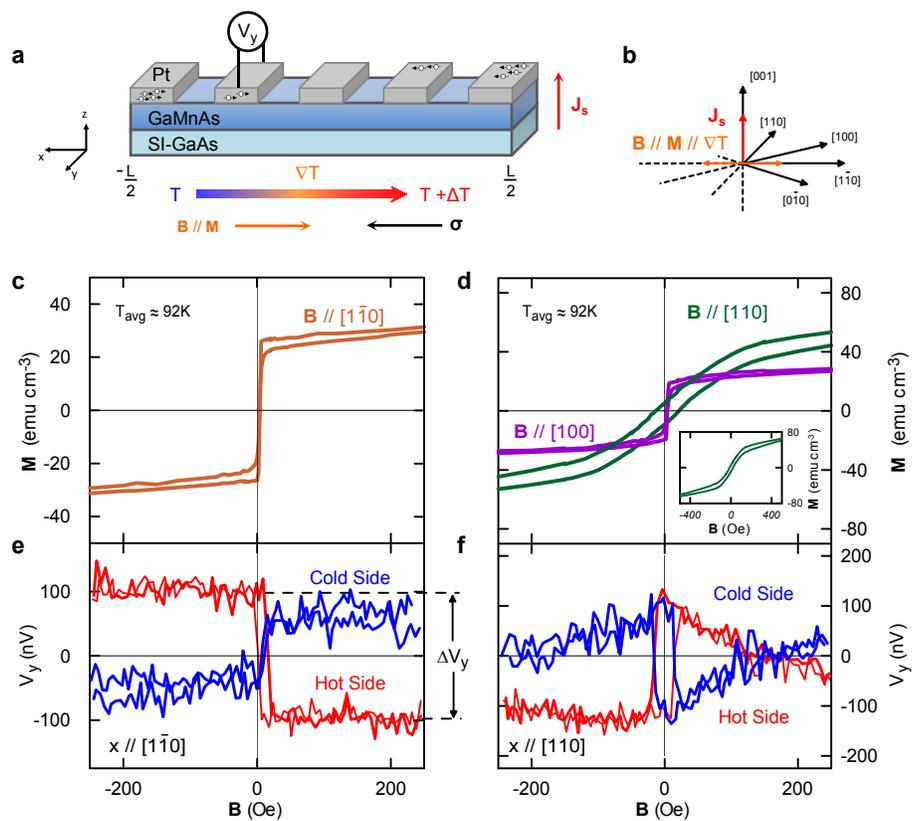

Figure 1, Jaworski, et. al.

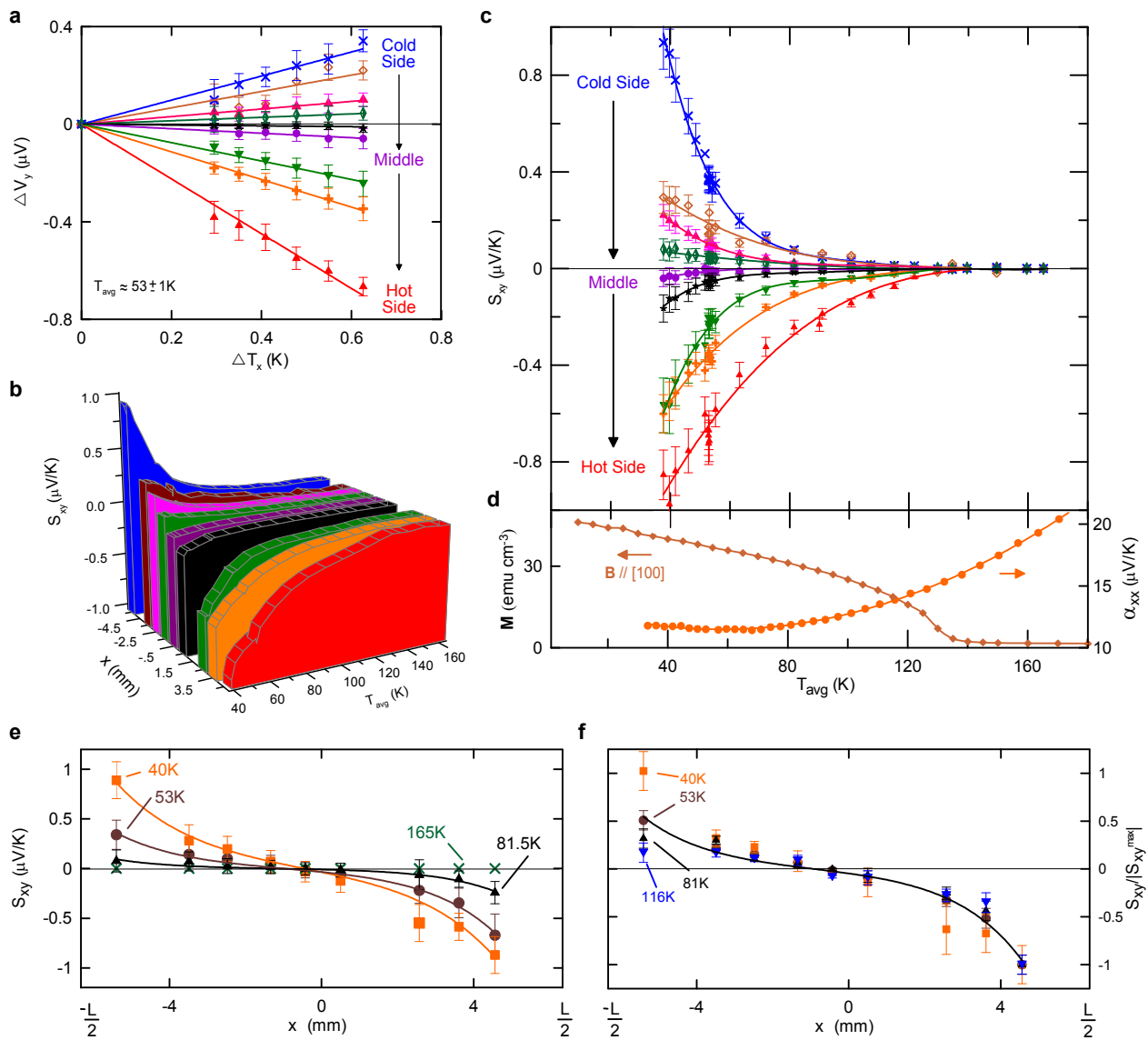

Figure 2, Jaworski, et. al.

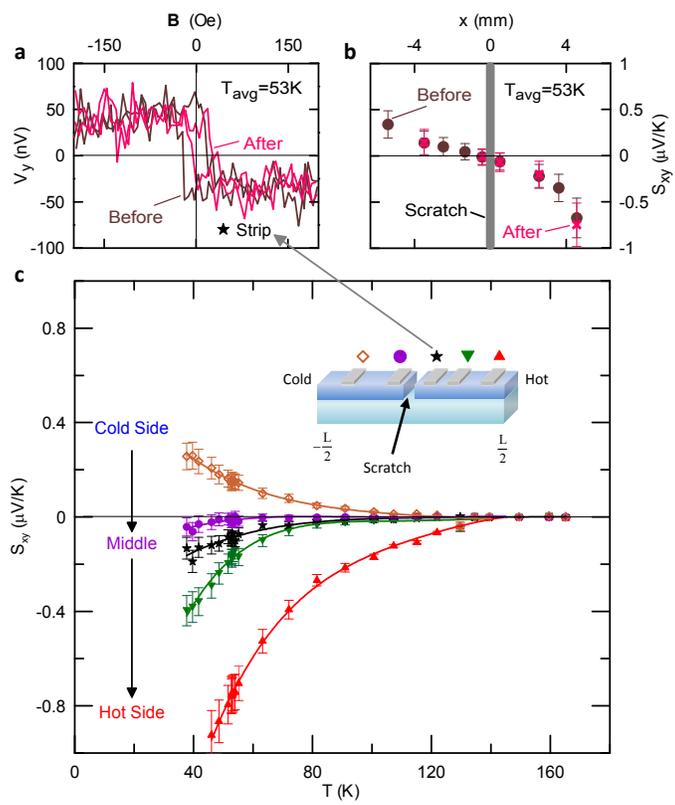

Figure 3, Jaworski, et. al.

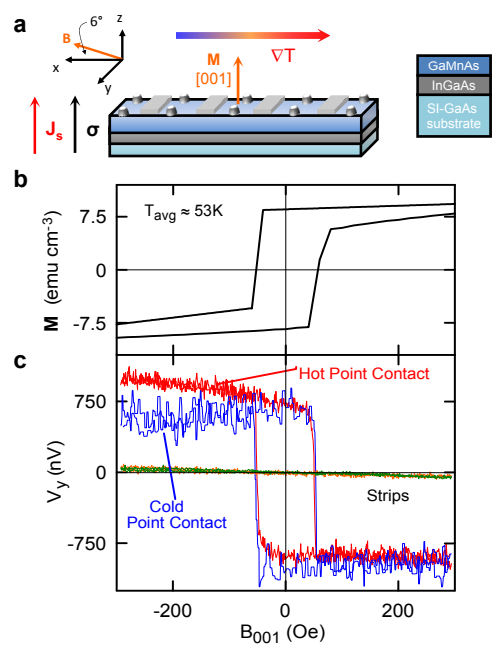

Figure 4, Jaworski, et. al.

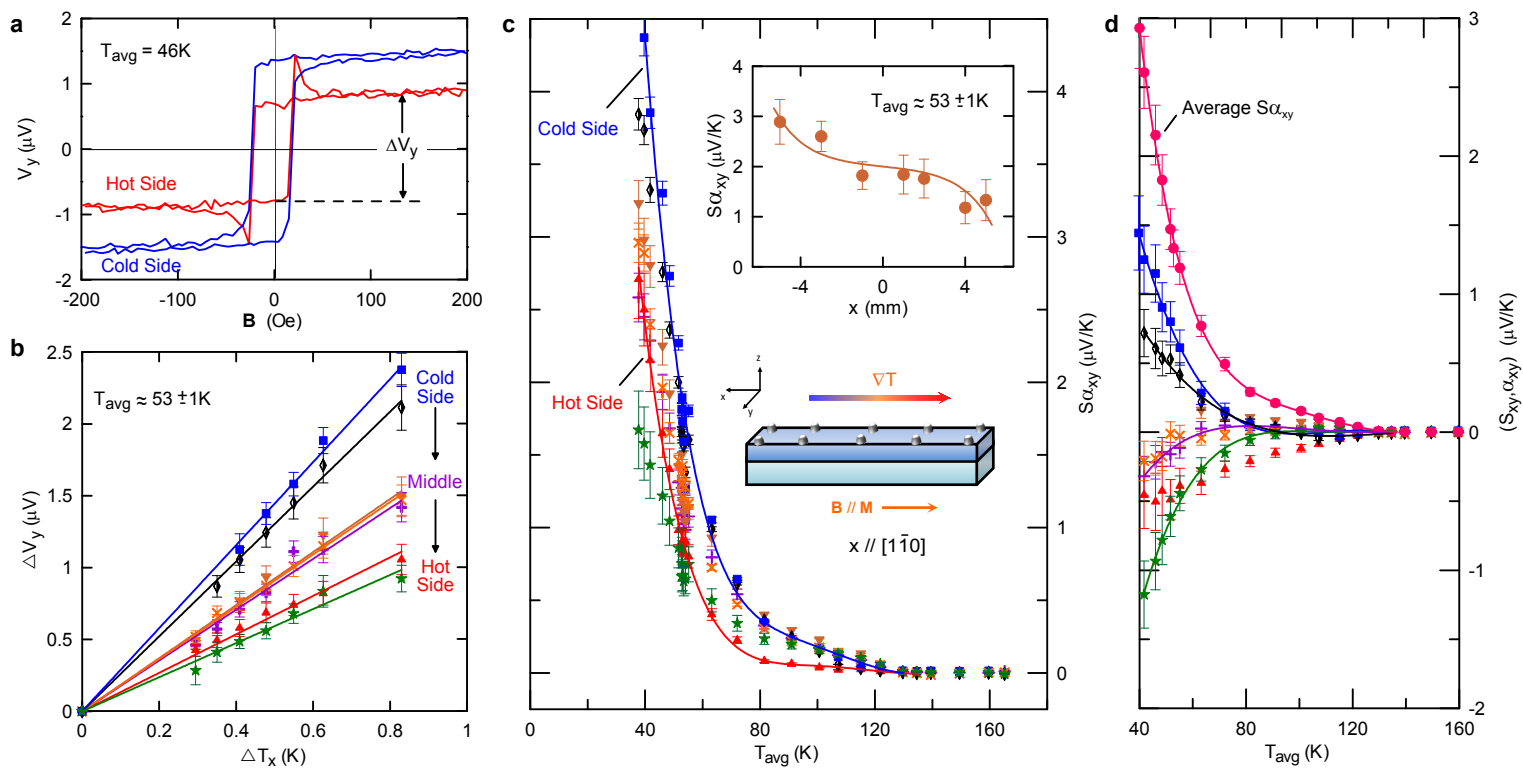

Figure 5, Jaworski, et. al.

Supplementary Information

# Observation of the Spin-Seebeck Effect in a Ferromagnetic Semiconductor

C. M. Jaworski[1,] J. Yang[2], S. Mack[3], D. D. Awschalom[3], J. P. Heremans[1,4*], R. C. Myers[2,4*]

1. Department of Mechanical Engineering, The Ohio State University, Columbus, OH
2. Department of Materials Science and Engineering, The Ohio State University, Columbus, OH
3. Center for Spintronics and Quantum Computation, University of California, Santa Barbara, CA
4. Department of Physics, The Ohio State University, Columbus, OH

*To whom correspondence should be addressed. Email: heremans.1@osu.edu, myers.1079@osu.edu

**Supplementary Methods**

The samples in this study are epitaxial $Ga_{1-s}Mn_sAs$ grown on [001] GaAs semi-insulating substrates using a Veeco Gen II MBE system. Wafers were grown using low temperature stoichiometric growth technique[1], in which the As:Ga ratio and therefore the stoichiometry of the films can be tuned[2]. These samples were grown at a substrate temperature of 150 °C. Growth rate calibration, substrate temperature monitoring, oxide desorption, high temperature GaAs buffer growth, cool down procedure, and *ex-situ* annealing are described in Ref. [1]. By stopping the substrate rotation, an As:Ga flux gradient across one direction [110] is generated. Stoichiometric graded samples were cleaved with the long axis only along the $[1\bar{1}0]$ direction, the direction along which the stoichiometry and %Mn is constant. In addition to the rotated (non-graded) samples discussed in the main text, we tested samples from three non-rotated wafers with s=0.138, 0.142, 0.16, and include additional data on s=0.16. An additional rotated wafer identical to the one in the main text (s=0.158) was prepared, followed by deposition of 10 nm of MnAs on top of the GaMnAs. This sample is then cleaved with the long axis along $[1\bar{1}0]$. Sample thicknesses are given in Table 1.

Refer to Fig. S1 for images of the measurement setup. Cernox thermometry is attached to gold plated manganin leads that are attached to the back of the wafer. This setup allows measurement of two transverse voltages while recording temperature gradients; after this initial run we remove the thermometry and use the free leads to record an additional three voltages. We use the same heater power and base temperature as the initial run, and use the corresponding $\Delta T_x$ in calculations. Removal of the thermometry and copper leads does not affect the signal. Furthermore, we have place the leads such that they extend from $+y$ or $-y$, again this does not change the signal. The small silver dots (12 in total, 6 on each side) along the edge of sample are the contacts where we silver epoxy copper wires. The sample chamber is enclosed by a gold-plated copper cap. The placement of the heater and sink on the sample is done to preclude a $\nabla_z T$. The sample is centered with respect to the alumina pads in the z-direction.

**Supplementary Discussion**

Figure S2 is a summary of the data from a non-rotated $Ga_{1-s}Mn_sAs$ sample with s=0.16. Refer to Fig. 1 of the main text for a schematic. $\nabla T_x$ is along $[1\bar{1}0]$ (stoichiometric direction) and $V_y$ is measured along [110]. Only platinum strips were tested, data on the bottom strip could not be obtained due to the sample cracking. Strip 1 is at the hot side, and strips 2 and 3 are mirrored around the center, with the temperature at strip 2 hotter than strip 3. Fig. S2c shows the raw voltage traces at each contact for differing $\Delta T_x$ values with an average sample temperature of 98K. Figure S2b plots $\Delta V_y$ vs. $\Delta T_x$ for the data in Fig. S2c and repeats the linear behavior. While not shown, $\Delta V_y$ vs. $\Delta T_x$ traces at other average sample temperatures also have the linear behavior and pass through the origin. $S_{xy}$ as a function of temperature is shown in Fig S2a. $S_{xy}$ is zero above $T_c$, and increases in magnitude with decreasing temperature, again repeating the temperature dependence of the sample in the main text. The positional dependence is also the same, thus reproducing the spin-Seebeck effect in a different sample.

We repeat the test for a macroscopic spin/charge current along *x* discussed in the main text and Fig. 3. by first shortening the sample ~15% and then polishing away a 0.35-mm wide region of GaMnAs with sandpaper in the middle of the sample, Fig. S3a (Case 2). After repeating the measurement, a second scratch is added just below the hot side Pt strip (Case 3). Fig. S3b shows hysteresis loops from the hot-side contact before and after each scratch, exhibiting no quantitative change. The temperature dependence of $S_{xy}$, is likewise unaffected by the presence of scratches, Fig. S3c. Further, $S_{xy}$ at the top edge of the sample shows no dependence on the overall length of the sample, confirming our normalization procedure. Obviously, changing the length of sample will change the midpoint of the sample, and thus affect the $\Delta V_y$ in these regions. This repeats the results shown in Fig. 3 of the main text on different GaMnAs sample confirming that the spin-Seebeck effect does not depend on longitudinal electrical communication.

We detail in Fig. S4 the temperature dependence of the Nernst-Ettingshausen effect as measured in point contacts on the tensile strained GaMnAs with s=0.056 and x // [110]. The Nernst Coefficient $\alpha_{xyz}$ is defined as $\alpha_{xyz} = \frac{E_y/B_z}{\nabla_x T}$ and a schematic is shown in the inset of Fig.

**Figure** S4. The magnetization of this sample lies out-of-plane, and thus the ISHE cross product is zero. To test this sample, we tilt it at approximately 6° from the xy-plane such that a component of the external magnetic field lies in the z-direction, thus allowing flipping of the magnetization out-of-plane (see inset). We performed magnetic field sweeps, as in Fig. 4 of the main text, at each temperature. Subtracting residuals, we calculate $\Delta V_y$, and after normalizing for $\nabla T_x$ and sample width, we obtain the Nernst Coefficient. This figure completes the data from Fig. 4 of the main text. This signal ($\Delta V_y$) is also linear in $\Delta T_x$, and has no spatial dependence, as expected. Again this signal goes to zero above $T_C$ as $\mathbf{M}_z = 0$, and increases in magnitude as the temperature is lowered. This measurement has increased

noise due to the lower carrier density of this sample compared with the other more highly Mn doped GaMnAs samples.

We have measured the Spin-Seebeck effect in a MnAs (10 nm)/GaMnAs (30 nm)/SI-GaAs ferromagnetic bilayer. Representative $V_y$ hysteresis loops are shown in Fig. S5 at 308K, which lies below the $T_C$ of MnAs (325 K) and above the $T_c$ of the GaMnAs. Interestingly, the sign is opposite that in GaMnAs. Full measurements over temperature (similar to Figs. 2 & 5) are currently underway, and will be published in a subsequent article. No steps in transverse voltage are observed either when the sample temperature is above the $T_C$ of MnAs, or when $\nabla T_x = 0$.

Table 1 lists all the samples measured in this study, and includes relevant sample properties. In summary, seven samples from 4 different wafers of GaMnAs with varying Mn concentration and **M** in-plane exhibit the spin-Seebeck signal.

**Figure Captions**

**Figure 1. Photographs of sample measurement apparatus**. The contacts are placed along the sample at the edges; this sample has 12 total, allowing the measurement of $V_y$ in 6 locations. The heater is located at the top of the sample, Alumina pads are used to electrically isolate the sample.

**Figure 2. Repeat of spin-Seebeck effect in a different GaMnAs sample**. Data in this figure was measured on GaMnAs with in-plane magnetization and s=0.16. **a,** $S_{xy}$ as a function of temperature for three Pt strips. Strip 1 is at the hot end, strips 2 and 3 are centered around the midpoint of the sample, with strip 3 on the cold half of the sample. **b,** $\Delta T_x$ vs. $\Delta V_y$ at $T_{avg}$~98K for the three contacts. This data is taken from the raw traces in (**c**). **c,** Raw traces with background voltages and EMF pickup subtracted for $\Delta T_x$~ 0, 3.5, 4.5, 5.9K for the three strips. Lines are added to guide the eye.

**Figure 3. Repeat of test for a longitudinal spin-current due to the spin-Seebeck effect and its dependence on temperature gradient.** All measurements taken with x // [1-10] on a sample cleaved from the same wafer as Figs. 1-3,5. **a,** Schematic for Case 1 (intact sample), Case 2 (sample was shortened 15% and a strip of GaMnAs was scratched away from the center of the sample), Case 3 (a second scratch was added to the sample). Each scratch was sufficient for complete electrical isolation (>3MΩ). **b,** $V_y$ was recorded during magnetic hysteresis loops from the hot-side contact before and after each scratch. $V_y$ was normalized by the temperature gradient, which changed in each case. **c, S**pin-Seebeck coefficient as a function of temperature for each case.

**Figure 4. Nernst coefficient versus temperature**. Data in this figure was measured on GaMnAs with out-of-plane magnetization and s=0.056 (same sample in Fig. 4). Point contact #1 is at the hot end, and the numbers monotonically increase toward the cold end. The insets include a schematic of the Nernst effect field and flux directions, as well as a schematic (not to scale) of the sample measured in this study. The externally applied magnetic field is approximately 6° off the xy-plane.

**Figure 5. Spin-Seebeck signal in MnAs.** $V_y$ hysteresis loops for hot and cold side Pt strips on MnAs with x / [1-10] with $\Delta T_x$ = 8.2K at $T_{avg}$ = 308K. $V_y$ switches at the coercive field of MnAs (not shown).

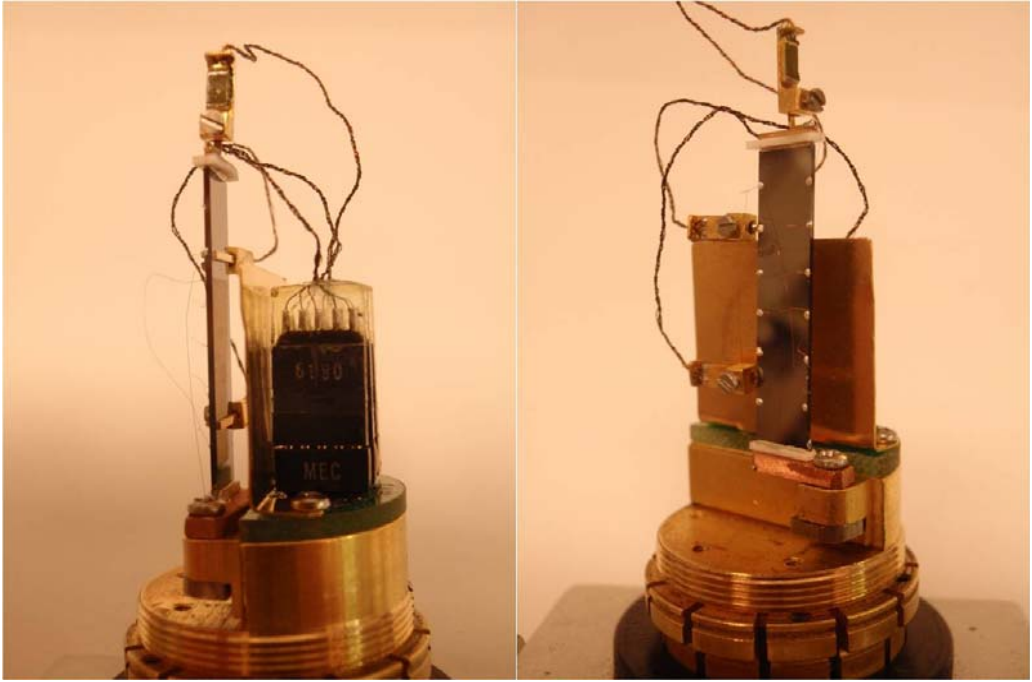

Figure 1

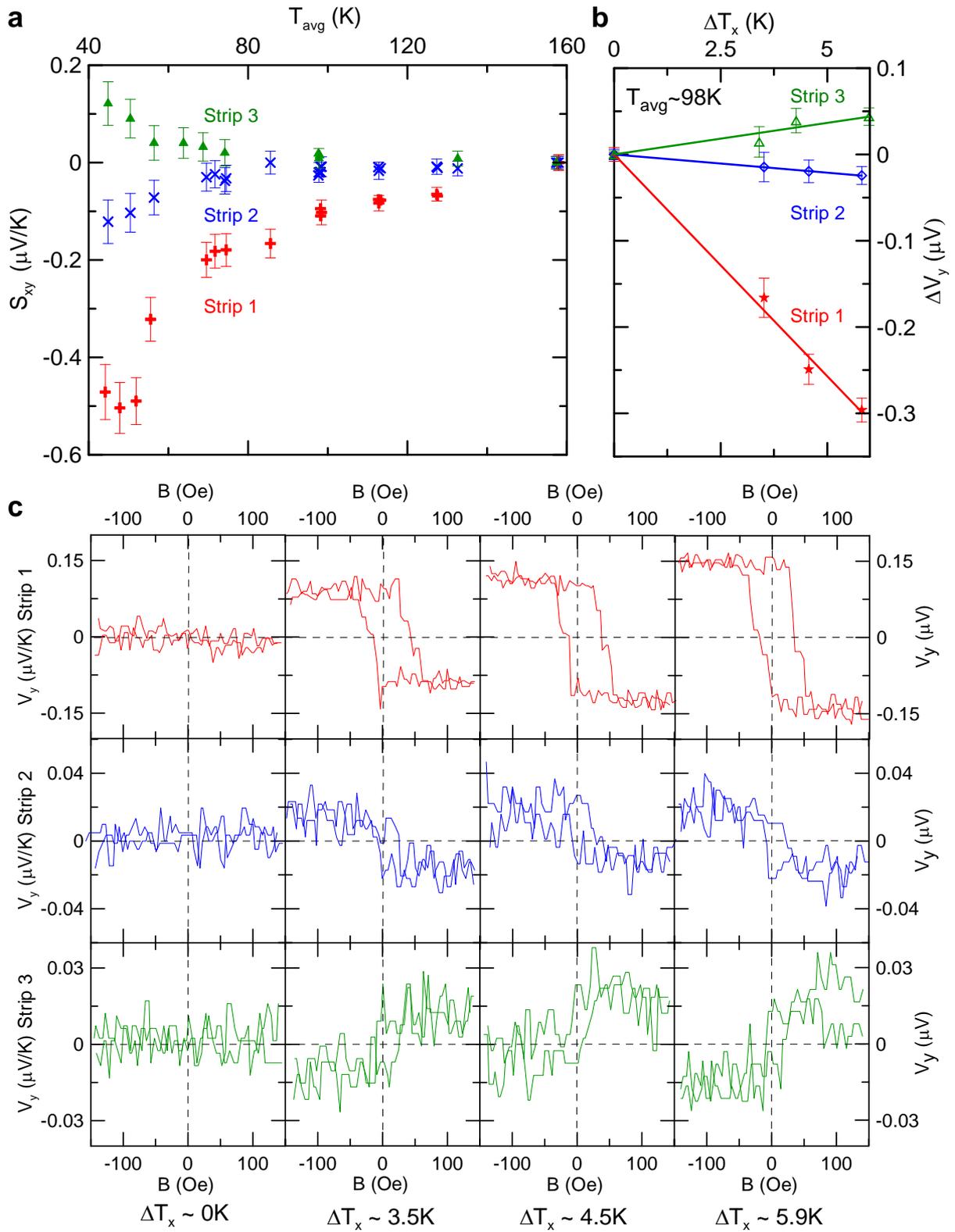

Figure 2

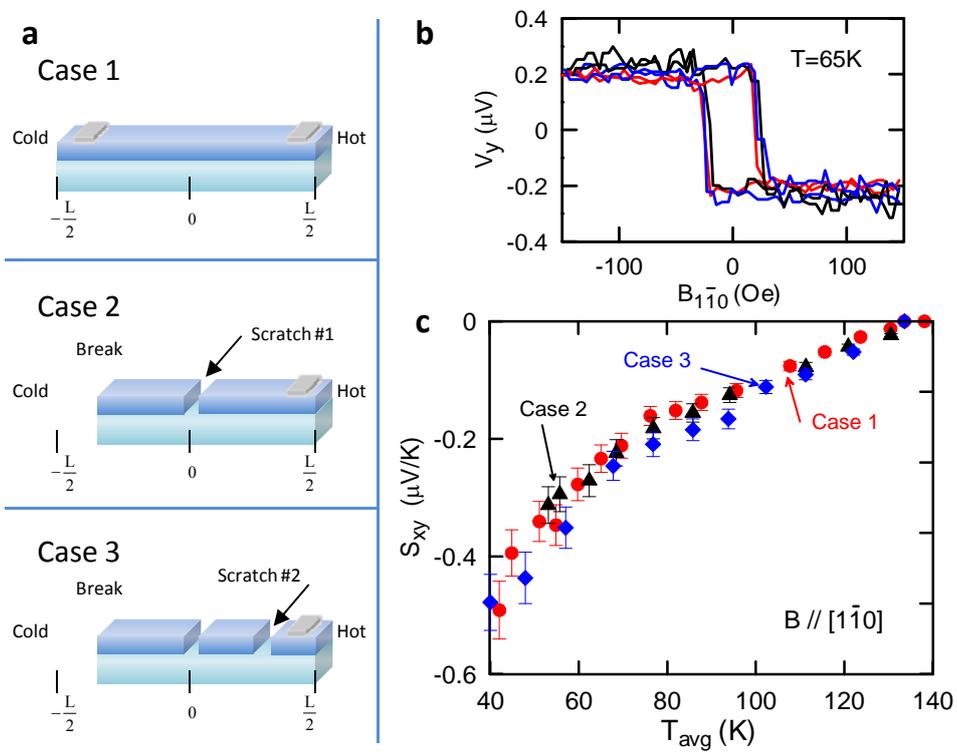

Figure 3

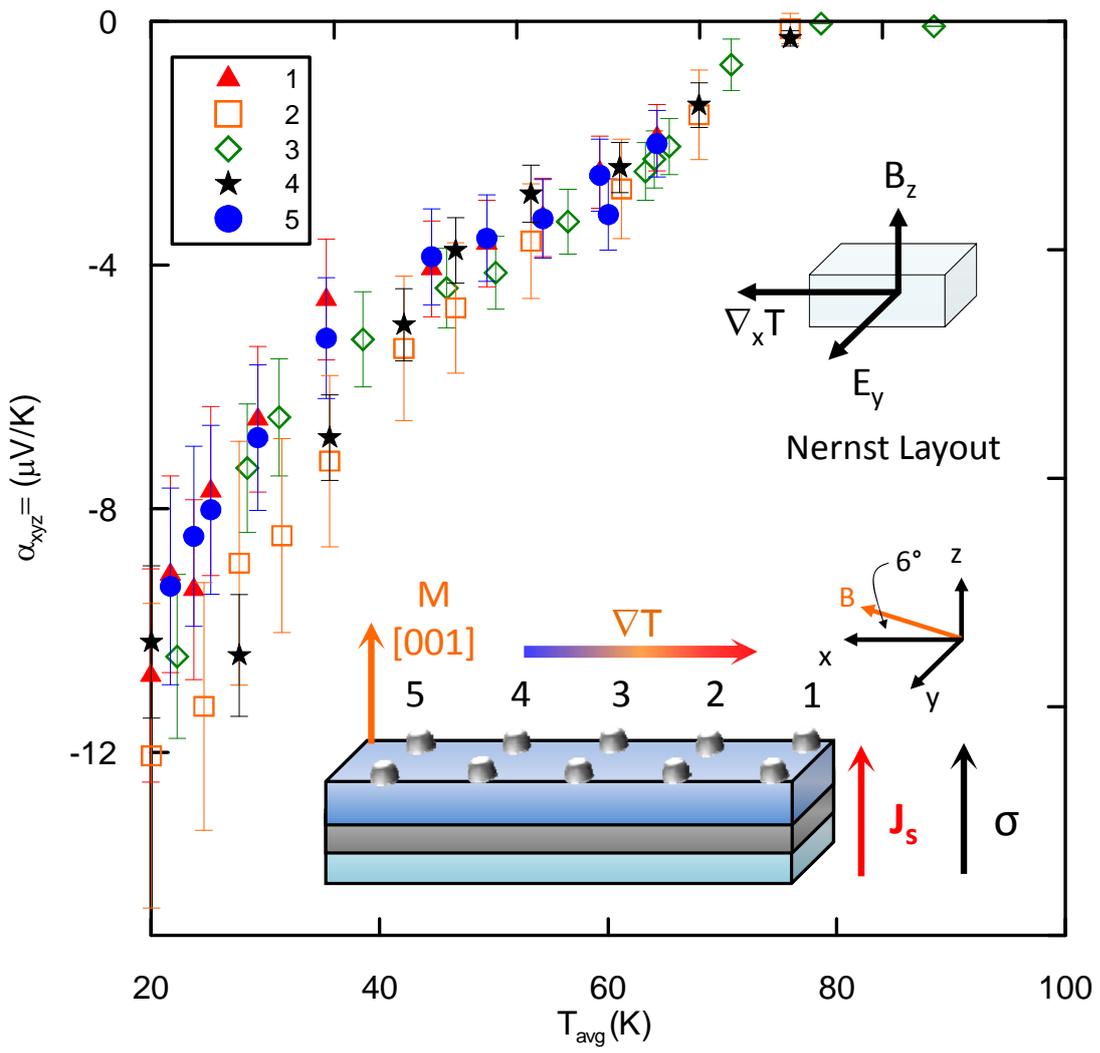

Figure 4

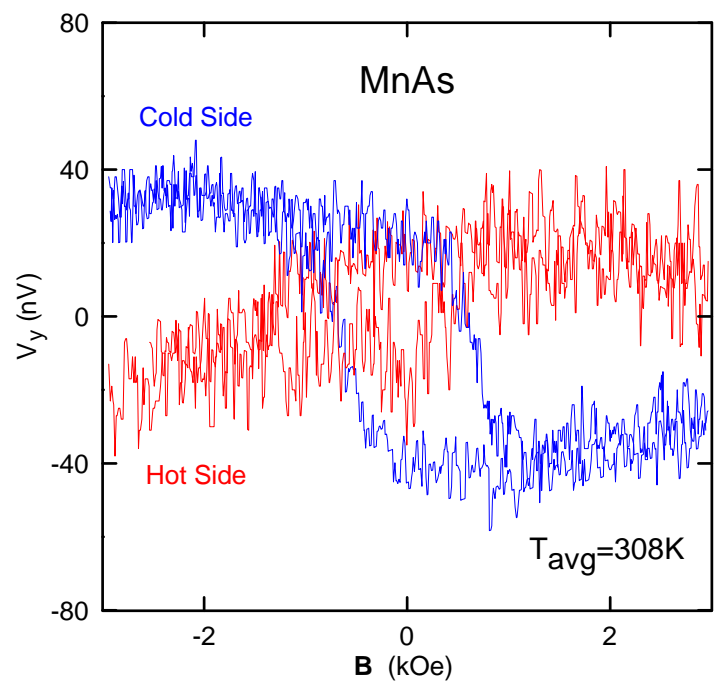

Figure 5

**Table 1**

| Sample # | x-direction | Rotated? | Thickness | Mn | Anneal | $M_{sat}$ at 50K | Tc | spin-Seebeck? | Data Presented? |
|---|---|---|---|---|---|---|---|---|---|
| | | | (nm) | (%) | (minutes) | (emu/cm^3) | (K) | | Figure # |
| 070606A | [1$\bar{1}$0] | No | 100 | 14.2 | no | 35 | 115 | Yes | No |
| 070313A | [1$\bar{1}$0] | No | 100 | 16 | yes | 40 | 160 | Yes | S2 |
| 061205B | [1$\bar{1}$0] | No | 100 | 13.8 | no | 20 | 115 | Yes | No |
| 080326B #1 | [110] | Yes | 30 | 15.8 | 30 (in-situ) | 80 | 140 | Yes | 1f |
| 080326B #2 | [1$\bar{1}$0] | Yes | 30 | 15.8 | 30 (in-situ) | 30 | 140 | Yes | 1-3,5 |
| 080326B #3 | [1$\bar{1}$0] | Yes | 30 | 15.8 | 30 (in-situ) | 30 | 140 | Yes | No |
| 080326B #4 | [1$\bar{1}$0] | Yes | 30 | 15.8 | 30 (in-situ) | 30 | 140 | Yes | S3 |
| 050519B | [1$\bar{1}$0] | Yes | 100 | 5.6 | no | 8 | 70 | No | 4, S4 |
| 080324B | [1$\bar{1}$0] | Yes | 10 | MnAs | no | - | 325 | Yes | S5 |